\documentclass{article}
\usepackage{amsmath,amsthm,amsfonts,amscd}
\DeclareMathOperator{\Hom}{Hom}
\DeclareMathOperator{\Com}{Com}
\DeclareMathOperator{\CCup}{Cup}
\DeclareMathOperator{\II}{I}
\DeclareMathOperator{\1}{id}
\DeclareMathOperator{\ad}{ad}
\DeclareMathOperator{\dev}{dev}
\DeclareMathOperator{\Ker}{Ker}
\DeclareMathOperator{\IM}{Im}
\newcommand{\NN}{\mathbb{N}}
\newcommand{\EEnd}{\mathcal End}
\newcommand{\EE}{\mathcal E}
\newcommand{\bul}{\bullet}
\newcommand{\de}{\delta}
\renewcommand{\u}{\smile}
\renewcommand{\=}{:=}
\renewcommand{\t}{\otimes}
\renewcommand{\o}{\circ}

\newtheorem{thm}{Theorem}[section]
\theoremstyle{definition}
 \newtheorem{defn}[thm]{Definition}
\theoremstyle{definition}
 \newtheorem{exam}[thm]{Example}
\begin{document}
\title{Operads for x-physics}
\date{}
\author{Eugen Paal\\ \\
Department of Mathematics, Tallinn Technical University\\
Ehitajate tee 5, 19086 Tallinn, Estonia\\
e-mail: eugen@edu.ttu.ee}
\maketitle
\thispagestyle{empty}
%
\begin{abstract}
The essential parts of the operad algebra are presented, which
should be useful when confronting with the operadic physics. It is also
clarified how the Gerstenhaber algebras can be associated with the linear
pre-operads (comp algebras). Their relation to mechanics is concisely
discussed. A hypothesis that the Feynman diagrams are observables
is proposed.

\end{abstract}

\section{Introduction and outline of the paper}

Operads, in essence, were invented by Gerstenhaber \cite{Ger,Ger68} and
Stasheff \cite{Sta63}. The notion of an operad was formalized by May
\cite{May72} as a tool for iterated loop spaces. In 1994/95
\cite{GeVo94,VoGe}, Gerstenhaber and Voronov published main principles of
the operad calculus. Quite a remarkable research activity on operad theory
and its applications can be observed in the last decade (e.~g.
\cite{Rene,Smi01}). It may be said that operads are
also becoming an interesting and important mathematical tool for QFT (e.~g.
\cite{KiStaVo95,KiVoZu97,Sta97,Sta99,Kre}) and deformation quantization
\cite{Kon}.

In this paper, the essential parts of the operad algebra are
presented, which should be useful when confronting with the operadic
physics. We start from simple algebraic axioms.
Basic algebraic constructions associated with a linear
pre-operad are introduced.
Their properties and the first derivation deviations of the
pre-coboundary operator are explicitly given. Under certain condition
(formal associativity constraint), the Gerstenhaber algebra structure
appears in the associated cohomology. At last, it is concisely discussed
how operads and Gerstenhaber algebras are related to mechanics.
A hypothesis that the Feynman diagrams are observables
is proposed.

\section{Pre-operad (composition system)}

Let $K$ be a unital associative commutative ring, and let $C^n$
($n\in\NN$) be unital $K$-modules. For \emph{homogeneous} $f\in
C^n$, we refer to $n$ as the \emph{degree} of $f$ and often write
(when it does not cause confusion) $f$ instead of $\deg f$. For
example, $(-1)^f\=(-1)^n$, $C^f\=C^n$ and $\o_f\=\o_n$. Also, it
is convenient to use the \emph{reduced} degree $|f|\=n-1$.
Throughout this paper, we assume that $\t\=\t_K$.

\begin{defn}
A linear \emph{pre-operad} (\emph{composition system}) with
coefficients in $K$ is a sequence $C\=\{C^n\}_{n\in\NN}$ of unital
$K$-modules (an $\NN$-graded $K$-module), such that the following
conditions hold.
\begin{enumerate}
\item[(1)]
For $0\leq i\leq m-1$ there exist \emph{partial compositions}
\[
  \o_i\in\Hom(C^m\t C^n,C^{m+n-1}),\qquad |\o_i|=0.
\]
\item[(2)]
For all $h\t f\t g\in C^h\t C^f\t C^g$,
the \emph{composition (associativity) relations} hold,
\[
(h\o_i f)\o_j g=
\begin{cases}
    (-1)^{|f||g|} (h\o_j g)\o_{i+|g|}f
                       &\text{if $0\leq j\leq i-1$},\\
    h\o_i(f\o_{j-i}g)  &\text{if $i\leq j\leq i+|f|$},\\
    (-1)^{|f||g|}(h\o_{j-|f|}g)\o_i f
                       &\text{if $i+f\leq j\leq|h|+|f|$}.
\end{cases}
\]
\item[(3)]
There exists a unit $\II\in C^1$ such that
\[
\II\o_0 f=f=f\o_i \II,\qquad 0\leq i\leq |f|.
\]
\end{enumerate}
\end{defn}

In the 2nd item, the \emph{first} and \emph{third} parts of the
defining relations turn out to be equivalent.

\begin{exam}[endomorphism pre-operad {\rm \cite{Ger,Ger68}}]
\label{HG} Let $A$ be a unital $K$-module and
$\EE_A^n\={\EEnd}_A^n\=\Hom(A^{\t n},A)$. Define the partial compositions
for $f\t g\in\EE_A^f\t\EE_A^g$ as
\[
f\o_i g\=(-1)^{i|g|}f\o(\1_A^{\t i}\t g\t\1_A^{\t(|f|-i)}),
         \qquad 0\leq i\leq |f|.
\]
Then $\EE_A\=\{\EE_A^n\}_{n\in\NN}$ is a pre-operad
(with the unit $\1_A\in\EE_A^1$) called the \emph{endomorphism pre-operad}
of $A$.
\end{exam}

\begin{exam}[associahedra]
A geometrical  example of a pre-operad is provided by the Stasheff
\emph{associahedra} \cite{Sta63}.
Quite a surprising realization of the associahedra as
\emph{truncated simplices} was discovered and studied in
\cite{ShnSte94,Sta97,Markl97}.
Markl and Shnider \cite{MarShn96a,MarShn96b,MarShn96c} revealed its relevance
to the Drinfel'd algebra \cite{Dri89} cohomology and deformations.
\end{exam}

\section{Associated operations}

Throughout this paper fix $\mu\in C^2$.

\begin{defn}
The \emph{cup-multiplication} $\u\:C^f\t C^g\to C^{f+g}$ is defined
by
\[
f\u g\=(-1)^f(\mu\o_0 f)\o_f g\in C^{f+g},
\qquad|\smile|=1.
\]
The pair $\CCup C\=\{C,\u\}$ is called a $\u$-algebra (cup-algebra) of $C$.
\end{defn}

\begin{exam}
For the endomorphism pre-operad (Example \ref{HG}) $\EE_A$ one has
\[
f\u g=(-1)^{fg}\mu\o(f\t g),
      \qquad \mu\t f\t g\in\EE_A^2\t\EE_A^f\t\EE_A^g.
\]
\end{exam}

\begin{defn}
The \emph{total composition} $\bul\:C^f\t C^g\to C^{f+|g|}$ is defined by
\[
f\bul g\=\sum_{i=0}^{|f|}f\o_i g\in C^{f+|g|},
     \qquad |\bul|=0.
\]
The pair $\Com C\=\{C,\bul\}$ is called the \emph{composition algebra} of $C$.
\end{defn}

\begin{defn}[tribraces and tetrabraces]
The Gerstenhaber \emph{tribraces} $\{\cdot,\cdot,\cdot\}$ are
defined as a double sum
\[
\{h,f,g\}\=\sum_{i=0}^{|h|-1}\sum_{i+f}^{|f|+|h|}(h\o_i f)\o_j g\in C^{h+|f|+|g|},
    \quad|\{\cdot,\cdot,\cdot\}|=0.
\]
The \emph{tetrabraces} $\{\cdot,\cdot,\cdot,\cdot\}$ are defined by
\[
\{h,f,g,b\}\=\sum_{i=0}^{|h|-2}\sum_{j=i+f}^{|h|+|f|-1}\sum_{k=j+g}^{|h|+|f|+|g|}
((h\o_{i}f)\o_{j}g)\o_{k}b\in C^{h+|f|+|g|+|b|}.
\]
\end{defn}

It turns out that $f\u g=(-1)^f\{\mu,f,g\}$. In general, $\CCup C$ is a
\emph{non-associative} algebra. By denoting $\mu^{2}\=\mu\bul\mu$ it turns
out that the associator in $\CCup C$ reads
\[
(f\smile g)\smile h-f\smile(g\smile h)=\{\mu^{2},f,g,h\}.
\]
Thus the \emph{formal associator} $\mu^{2}$ is an
obstruction to associativity of $\CCup C$. For the endomorphism pre-operad
$\EE_A$, $\mu^{2}$ reads as an associator as well:
\[
\mu^{2}=\mu\o(\mu\t\1_A-\1_A\t\mu),\qquad\mu\in\EE_A^2.
\]

\section{Identities}

In a pre-operad $C$, the Getzler identity
\[
(h,f,g) \=(h\bul f)\bul g-h\bul(f\bul g)
         =\{h,f,g\}+(-1)^{|f||g|}\{h,g,f\}
\]
holds, which easily implies the Gerstenhaber identity
\[
(h,f,g)=(-1)^{|f||g|}(h,g,f).
\]
The \emph{commutator} $[\cdot,\cdot]$ is defined in $\Com C$ by
\[
[f,g]\=f\bul g-(-1)^{|f||g|}g\bul f=-(-1)^{|f||g|}[g,f],\qquad|[\cdot,\cdot]|=0.
\]
The \emph{commutator algebra} of $\Com C$ is denoted as
$\Com^{-}\!C\=\{C,[\cdot,\cdot]\}$. By using the Gerstenhaber identity, one
can prove that $\Com^-\!C$ is a \emph{graded Lie algebra}. The Jacobi
identity reads
\[
(-1)^{|f||h|}[[f,g],h]+(-1)^{|g||f|}[[g,h],f]+(-1)^{|h||g|}[[h,f],g]=0.
\]

\section{Pre-coboundary operator}

In a pre-operad $C$, define a \emph{pre-coboundary} operator $\de\=\de_\mu$ by
\begin{align*}
-\de f&\=\ad_\mu^{right}f\=[f,\mu]\=f\bul\mu-(-1)^{|f|}\mu\bul f \\
      &\,\,=f\u\II+f\bul\mu+(-1)^{|f|}\,\II\u f,\qquad \deg\de=+1=|\de|.
\end{align*}
It turns out that $\de^{2}_\mu=-\de_{\mu^{2}}$. It follows from the Jacobi
identity in $\Com^{-}\!C$ that $\de$ is a (right) derivation of
$\Com^{-}\!C$,
\[
\de[f,g]=(-1)^{|g|}[\de f,g]+[f,\de g].
\]
But $\delta$ need not be a derivation of $\CCup C$, and $\mu^{2}$ again appears
as an obstruction:
\[
\de(f\smile g)-f\smile\delta g-(-1)^{g}\delta f\smile g
=(-1)^{|g|}\{\mu^{2},f,g\}.
\]

\section{Derivation deviations}

The \emph{derivation deviation} of $\de$ over $\bul$ is defined by
\[
\dev_\bul\de(f\t g)
   \=\de(f\bul g)-f\bul\de g-(-1)^{|g|}\de f\bul g.
\]

\begin{thm}
\label{first}
In a pre-operad $C$, one has
\[
(-1)^{|g|}\dev_\bul\de(f\t g)=f\u g-(-1)^{fg}g\u f.
\]
\end{thm}
The derivation deviation of $\de$ over $\{\cdot,\cdot,\cdot\}$ is defined by
\begin{align*}
\dev_{\{\cdot,\cdot,\cdot\}}\de\,(h\t f\t g)
    \=\de\{h,f,g\}
     &-\{h,f,\de g\}\\
     &-(-1)^{|g|}\{h,\de f,g\}-(-1)^{|g|+|f|}\{\de h,f,g\}.
\end{align*}

\begin{thm}
\label{second}
In a pre-operad $C$, one has
\[
(-1)^{|g|}\dev_{\{\cdot,\cdot,\cdot\}}\de\,(h\t f\t g)=
    (h\bul f)\u g+(-1)^{|h|f}f\u(h\bul g)-h\bul(f\u g).
\]
\end{thm}
Thus the \emph{left} translations in $\Com C$ are not
derivations of $\CCup C$, the corresponding deviations are related to
$\dev_{\{\cdot,\cdot,\cdot\}}\de$. It turns out that the \emph{right}
translations in $\Com C$ are derivations of $\CCup C$,
\[
(f\u g)\bul h=f\u(g\bul h)+(-1)^{|h|g}(f\bul h)\u g.
\]
By combining this formula with the one from Theorem \ref{second} we obtain

\begin{thm}
\label{second*}
In a pre-operad $C$, one has
\[
(-1)^{|g|}\dev_{\{\cdot,\cdot,\cdot\}}\de\,(h\t f\t g)=
    [h,f]\u g+(-1)^{|h|f}f\u[h,g]-[h,f\u g].
\]
\end{thm}

\section{Associated cohomology and Gerstenhaber algebra}

Now, it can be clarified how the Gerstenhaber algebra can be associated with a
linear pre-operad. If (formal associativity) $\mu^{2}=0$ holds, then
$\de^{2}=0$, which in turn implies $\IM\de\subseteq\Ker\de$. Then one can
form an associated cohomology ($\NN$-graded module) $H(C)\=\Ker\de/\IM\de$ with
homogeneous components
\[
H^{n}(C)\=\Ker(C^{n}\stackrel{\de}{\rightarrow}C^{n+1})/
           \IM(C^{n-1}\stackrel{\de}{\rightarrow}C^{n}),
\]
where, by convention, $\IM(C^{-1}\stackrel{\de}{\rightarrow}C^{0})\=0$.
Also, in this ($\mu^{2}=0$) case, $\CCup C$ is \emph{associative} and $\de$
is a \emph{derivation} of $\CCup C$. Recall from above that $\Com^{-}\!C$
is a graded Lie algebra and $\de$ is a derivation of $\Com^{-}\!C$. Due to
the derivation properties of $\de$, the multiplications $[\cdot,\cdot]$ and
$\smile$ induce corresponding (factor) multiplications on $H(C)$, which we
denote by the same symbols. Then $\{H(C),[\cdot,\cdot]\}$ is a \emph{graded
Lie algebra}. It follows from Theorem \ref{first} that the induced
$\smile$-multiplication on $H(C)$ is \emph{graded commutative},
\[
f\smile g=(-1)^{fg}g\smile f
\]
for all $f\t g\in H^{f}(C)\t H^{g}(C)$, hence $\{H(C),\smile\}$ is an
\emph{associative graded commutative} algebra.
It follows from Theorem \ref{second*} that the  \emph{graded Leibniz rule}
holds,
\[
[h,f\u g]=[h,f]\u g+(-1)^{|h|f}f\u[h,g]
\]
for all $h\t f\t g\in H^{h}(C)\t H^{f}(C)\t H^{g}(C)$. At last, it is also
relevant to note that
\[
0=|[\cdot,\cdot]|\neq|\smile|=1.
\]
In this way, the triple $\{H(C),\smile,[\cdot,\cdot]\}$ turns out to be a
\emph{Gerstenhaber algebra}. This fact is a known from
\cite{Ger68,GGS92Am,CGS93}.

In the case of an endomorphism pre-operad, the Gerstenhaber algebra
structure appears on the Hochschild cohomology of an associative algebra
\cite{Ger}.

\section{Discussion: x-mechanics}

Some people like commutative diagrams. Consider the following one:
\[
\begin{CD}
\text{Poisson algebras}@<\text{algebraic abstraction}<<\text{mechanics}\\
      \wr                       @.                        \wr\\
\text{Gerstenhaber algebras}@<\text{algebraic abstraction}<<\text{x-mechanics}
\end{CD}
\]
\emph{Poisson algebras} can be seen as an algebraic abstraction
of mechanics.
Here $\sim$ means \emph{similarity}:
Gerstenhaber algebras are graded analogs of the Poisson algebras.

It may be expected that there exists a kind of mechanics (x-mechanics)
associated with operads and Gerstenhaber algebras.
According to the diagram, x-mechanics is a graded analogue of mechanics and
\emph{observables} of an x-mechanical model must satisfy the
(homotopy \cite{GeVo94,VoGe}) Gerstenhaber algebra identities.

Recently Kreimer \cite{Kre} explained how the \emph{insertion operad of
Feynman graphs} is present in renormalization in QFT because insertion
operations are used in the Hopf algebra of Feynman graphs. Thus from the
operad theoretical point of view it may be expected that the
\emph{Feynman diagrams can be seen as observables}.

\section*{Acknowledgement}
Research supported by the ESF grant 3654.

\end{document}